\documentclass[a4paper,11pt]{article}

\usepackage{jheppub} 

\usepackage{graphicx,bm,float}
\usepackage[normalem]{ulem}

\makeatletter

\title{Two-dimensional lattice SU($N_c$) gauge theories 
  with multiflavor adjoint scalar fields}

\author[a]{Claudio Bonati} 
\author[a]{Alessio Franchi} 
\author[b]{Andrea Pelissetto}
\author[a]{Ettore Vicari} 

\affiliation[a]{Dipartimento di Fisica dell'Universit\`a di Pisa and INFN, Pisa, Italy}
\affiliation[b]{Dipartimento di Fisica dell'Universit\`a di Roma Sapienza and INFN, Roma, Italy}

\emailAdd{claudio.bonati@unipi.it}
\emailAdd{alessio.franchi@phd.unipi.it}
\emailAdd{Andrea.pelissetto@roma1.infn.it}
\emailAdd{ettore.vicari@unipi.it}

\date{\today}

\abstract{ We consider two-dimensional lattice SU($N_c$) gauge
  theories with $N_f$ real scalar fields transforming in the adjoint
  representation of the gauge group and with a global O($N_f$)
  invariance. Focusing on systems with $N_f\ge 3$, we study their
  zero-temperature limit, to understand under which conditions a
  continuum limit exists, and to investigate the nature of the
  associated quantum field theory. Extending previous analyses, we
  address the role that the gauge-group representation and the quartic
  scalar potential play in determining the nature of the continuum
  limit (when it exists). Our results further corroborate the
  conjecture that the continuum limit of two-dimensional lattice gauge
  models with multiflavor scalar fields, when it exists, is associated
  with a $\sigma$ model defined on a symmetric space that has the same
  global symmetry as the lattice model.  }

\begin{document}
\maketitle
\flushbottom


\section{Introduction}
\label{intro}

Gauge theories represent a unifying theme of modern theoretical
physics, being used to describe both fundamental processes in
high-energy particle theories~\cite{Weinberg-book, Wilson-74, ZJ-book}
and emerging phenomena in condensed matter physics~\cite{ZJ-book,
  Sachdev-19, Anderson-book}.  In the framework of statistical field
theory, one is typically interested in determining the low-energy
spectrum of the theory, the phase structure (in the context of gauge
theories with scalar fields the different phases are related to
different realizations of the Higgs mechanism), and the nature of
their critical behavior, or equivalently their continuum limit.  A
deep understanding of the interplay between the global and the local
symmetries of the theory is of fundamental importance for all these
topics.  In this paper we address such a problem in two-dimensional
(2D) lattice gauge theories, to identify the key features that
eventually determine the nature of their continuum limit and critical
behavior.

According to the Mermin-Wagner theorem \cite{MW-66, M-67}, 2D models with
global continuous symmetries do not show magnetized phases characterized by the
condensation of an order parameter, and therefore they do not undergo phase
transitions associated with the spontaneous breaking of the global symmetry.
However, 2D systems with global nonabelian symmetries may develop a critical
behavior in the zero-temperature limit.  For example, in the O($N$) $\sigma$
model with $N\ge 3$ and in the CP$^{N-1}$ model with $N\ge 2$, correlation
functions in the thermodynamic limit are characterized by a length scale $\xi$
that diverges as  $T^p e^{c/T}$ for $T\to 0$; see, e.g.,
Refs.~\cite{ZJ-book,PV-02}.  Systems with an Abelian O(2) global symmetry are
peculiar in this respect, since they may undergo a finite-temperature
topological Berezinskii-Kosterlitz-Thouless (BKT)
transition~\cite{KT-73,Berezinskii-70,Kosterlitz-74}, which separates the
high-$T$ disordered phase from the low-temperature nonmagnetized spin-wave
phase characterized by correlation functions that decay algebraically.

In the case of models characterized by both global and gauge symmetries, the
asymptotic critical behavior is expected to arise from the interplay between
the two different symmetries. For the purpose of understanding which features
are relevant and which  continuum limits are effectively realized, several 2D
lattice models presenting both global and gauge continuous symmetries have been
investigated~\cite{BPV-19-ah2,BPV-20-qcd2,BFPV-20-ong,BFPV-21}, such as the
lattice Abelian-Higgs model characterized by a global SU($N_f$) ($N_f\ge 2$)
and a local U(1) symmetry, the lattice scalar quantum chromodynamics with a
global SU($N_f$) and a local SU($N_c$) symmetry, and a lattice SO($N_c$) gauge
model with a global O($N_f$) and a local SO($N_c$) symmetry.  These studies
support the following general conjecture: the universal low-temperature
critical behavior, and therefore the continuum limit, of 2D lattice gauge
models with scalar fields is the same as that of 2D $\sigma$ models defined on
symmetric spaces~\cite{BHZ-80,ZJ-book}, which have  the same global symmetry.

In this paper we extend the above analyses in two different
directions.  First, we want to understand whether the above conjecture
also holds when the matter fields transform under a higher (than the
fundamental) representation of the nonabelian gauge group. Second, we
consider general quartic potentials, that allow us to obtain different
low-temperature behaviors.  For this purpose we consider a 2D lattice
gauge model with a matrix scalar field, which is invariant under
O($N_f$) global transformations and SU($N_c$) gauge transformations,
and in which the scalar field transforms according to the adjoint
representation of the gauge group.  It is worth mentioning that, for
$N_c = 2$, this model has been recently considered as an emerging
gauge theory for high-$T_c$ superconductors~\cite{SSST-19,SPSS-20}.

The above issues are investigated by scrutinizing the nature of the
low-energy configurations that are relevant in the zero-temperature
limit, and by performing numerical finite-size scaling (FSS) analyses
of Monte Carlo (MC) results. We present results for $N_c=2,\,3$ and
$N_f=3,\,4$. As we shall see, our results confirm the aforementioned
conjecture.  We consider first a scalar model which is maximally
symmetric in the absence of the gauge fields, i.e., it is an O($M$)
$\sigma$ model with $M=N_f \,(N_c^2-1)$. In this case the lattice
gauge model with $N_f$ scalar flavors in the adjoint gauge-group
representation shows an asymptotic zero-temperature critical behavior
that belongs to the universality class of the 2D RP$^{N_f-1}$ model,
defined on the symmetric space O($N_f$)/O($N_f-1$).  Then, we
generalize the model introducing a scalar potential that reduces the
symmetry of the ungauged model to O($N_f$)$\otimes$O($N_c^2-1$). In
this case, different behaviors are observed, depending on the sign of
one of the parameters appearing in the quartic potential. For negative
values of the parameter, the RP$^{N_f-1}$ behavior is still
observed. A different behavior is observed instead for positive
values. If $N_f\le N_c^2-1$, no continuum limit can be defined:
correlations are always short-ranged, even in the zero-temperature
limit. On the other hand, for $N_f > N_c^2-1$, long-range correlations
are observed. We conjecture that the continuum limit is associated
with a $\sigma$ model defined in the symmetric space
O($N_f$)/O($q$)$\otimes$O($N_f-q$) with $q=N_c^2-1$.  Numerical
results for $N_c=2$ and $N_f=4$ are in full agreement with this
conjecture.  The different behavior for positive and negative values
of the quartic potential parameter is due to the qualitative
differences of the minimum-action configurations that control the
zero-temperature limit.

The paper is organized as follows. In Sec.~\ref{model} we define the
lattice SU($N_c$) gauge model with scalar fields in the adjoint
representation.  In Sec.~\ref{varsce} we discuss the expected
low-temperature behavior. We determine the minimum-action
configurations and derive the corresponding effective models. In
Sec.~\ref{numres} we present Monte Carlo results that fully confirm
the predictions of Sec.~\ref{varsce}.  Finally, in Sec.~\ref{conclu}
we summarize our results and draw our conclusions.  In App.~\ref{AppA}
we study the role that gauge fields play in determining the relevant
low-temperature configurations.  In App.~\ref{MCsim} we report some
details of the MC simulations.

\section{2D lattice SU($N_c$) gauge models with scalar fields in the adjoint 
SU($N_c$) representation}
\label{model}

We consider multiflavor lattice gauge models defined on a square
lattice of linear size $L$ with periodic boundary conditions, which are
invariant under local SU($N_c$) and global O($N_f$) transformations. 
The fundamental variables are real matrices $\Phi^{af}_{\bm x}$ defined on the
sites of the lattice, with $a=1,...,N_c^2-1$ ({\em color} index) and
$f=1,...,N_f$ ({\em flavor} index).
They transform under the adjoint representation of the 
SU($N_c$) gauge group and under the fundamental representation of the 
O($N_f$) group: 
\begin{equation}
\Phi^{af}_{\bm x} = \sum_{b} \widetilde{V}_x^{ab} \Phi^{bf}_{\bm x} \qquad\qquad
\Phi^{af}_{\bm x} = \sum_{b} W^{fg} \Phi^{ag}_{\bm x}, 
\label{transformPhi}
\end{equation}
where $\widetilde{V}_x$ is a matrix belonging to the adjoint
representation of the SU($N_c$) gauge group and $W$ is an orthogonal matrix.
Using the Wilson approach~\cite{Wilson-74}, 
we introduce gauge variables $U_{{\bm x},\mu}
\in {\rm SU}(N_c)$ associated with each link $({\bm x},\mu)$ 
of the lattice. The model is
defined by the partition function
\begin{eqnarray}
  Z = \sum_{\{\Phi,U\}} e^{-\beta S}\,,\qquad \beta=1/T\,,
\qquad   S  = S_K(\Phi,U) + S_V(\Phi) + S_G(U)\,,
\label{hgauge}
\end{eqnarray}
where the action $S$ is written as a sum of three terms: $S_K$ is the 
kinetic term for the scalar field, $S_V$ is the local scalar potential,
and $S_G$ is the gauge action.

The kinetic term $S_K$ is given by
\begin{eqnarray}
  S_K(\Phi,U) =   - J
  {N_f\over 2} \sum_{{\bm x},\mu} {\rm Tr} \,\Phi_{\bm x}^t \,
  \widetilde{U}_{{\bm x},\mu} \, \Phi_{{\bm x}+\hat{\mu}}^{\phantom t}\,,
  \label{Kinterm}
\end{eqnarray}
where $\widetilde{U}^{ab}_{{\bm x},\mu}$ is the adjoint
representation of the link variable $U_{{\bm x},\mu}$. It can be written as 
\begin{equation}
\widetilde{U}^{ab} = 2 \, {\rm Tr}\,U^\dagger T^a U T^b\,,\qquad
a,b=1,...,N_c^2-1\,,
\label{utilde}
\end{equation}
where $T^a$ are the $(N_c^2-1)$ generators of the SU($N_c$) algebra 
in the fundamental representation,
normalized so that ${\rm Tr} \, T^a T^b =
\frac{1}{2}\delta^{ab}$.~\footnote{Using the completeness relation 
$\sum_{a} T_{ij}^a T_{kl}^a = \frac{1}{2} (\delta_{il}\delta_{jk} - N_c^{-1}
\delta_{ij}\delta_{kl})$ it is easily shown that
$\widetilde{U}^{ab}$ is a representation of SU($N_c$). Close to the identity, 
if $U_{ij}\approx \delta_{ij} + i \varepsilon^a T^a_{ij}$, one obtains
$\widetilde{U}^{ab}\approx \delta^{ab} + i \varepsilon^c (-i f^{abc})$, where
$f^{abc}$ are the structure constants of the SU($N_c$) group satisfying 
 $[T^a,T^b] = i f^{abc}T^c$. This proves that $\widetilde{U}$ belongs to
the adjoint representation.} 
We set the lattice spacing equal to one, so that all lengths are
measured in units of the lattice spacing. Using Eq.~(\ref{utilde}) we can
rewrite the kinetic term as 
\begin{eqnarray}
  S_K =   - J N_f \sum_{{\bm x},\mu}
  \sum_f {\rm Tr} \,U_{{\bm x},\mu}^\dagger \phi^f_{\bm x} \,
  U_{{\bm x},\mu} \, \phi^f_{{\bm x}+\hat{\mu}}\,,
  \label{Kinterm2}
\end{eqnarray}
where the trace is taken in the fundamental represention of SU($N_c$) and 
\begin{equation}
\phi^f_{ij} = \sum_a \Phi^{af} T^a_{ij}\,,\qquad 
\Phi^{af} = 2 \, {\rm Tr} \,\phi^f T^a\, . 
\label{psidef}
\end{equation}
In the following we set $J=1$, so that energies are measured in
units of $J$. 

The scalar potential term $S_V$ can be written as
\footnote{One can easily express the potential in terms of the 
variable $\phi$ defined in Eq.~(\ref{psidef}) using
${\rm Tr}\, \Phi^t \Phi = 2 \sum_f {\rm Tr}\, \phi^f \phi^f$ and
${\rm Tr} \,\Phi^t \Phi \Phi^t \Phi = 4 \sum_{fg} ({\rm Tr} \, \phi^f
\phi^g)^2$.}
\begin{eqnarray}
  S_V(\Phi) = \sum_{\bm x} V(\Phi_{\bm x})\,,\qquad 
  V(\Phi)={r\over 2} \, {\rm Tr}\,\Phi^t\Phi + {u\over 4} \, \left( {\rm
  Tr}\,\Phi^t\Phi\right)^2 + {v\over 4} \, {\rm Tr}\,\Phi^t\Phi
\Phi^t\Phi\,,\qquad
  \label{potential}
\end{eqnarray}  
which is the most general quartic potential that is invariant under 
O($N_f$)$\otimes$O($N_c^2-1$) transformations.
Note that, for $v=0$, the symmetry of the scalar potential enlarges
to O($M$) with $M=N_f(N_c^2-1)$. 
Finally, we define the gauge action
\begin{eqnarray}
S_G(U) =
- {\gamma\over N_c} \sum_{{\bm x}}  {\rm Re} \, {\rm Tr}\, \Pi_{\bm x}\,,
\qquad \Pi_{\bm x}=
      U_{{\bm x},1} \,U_{{\bm x}+\hat{1},2} \,U_{{\bm
    x}+\hat{2},1}^\dagger \,U_{{\bm x},2}^\dagger \,,\qquad
\label{plaquette}
\end{eqnarray}
in which the plaquette parameter $\gamma$ plays the role of inverse gauge
coupling. 

The action $S$ defined in Eq.~(\ref{hgauge}) is invariant under the
global O($N_f$) transformations (\ref{transformPhi}) and under local
SU($N_c$) transformations (the scalar field transforms as in
Eq.~(\ref{transformPhi}), while $U_{{\bm x},\mu} \to V_{\bm x} U_{{\bm
    x},\mu} V_{{\bm x}+\hat{\mu}}^\dagger$; $\widetilde{V}$
corresponds to $V$ in the adjoint representation).  For
$\gamma\to\infty$ the link variables $U_{{\bm x},\mu}$ become equal to
the identity, modulo gauge transformations. Thus, in this limit, one
recovers a matrix scalar model. For $v\not=0$ the global symmetry
group of this scalar model is O($N_f$)$\otimes$O($N_c^2-1$). For $v=0$
the symmetry group is O($M$) with $M=N_f(N_c^2-1)$.

For $\gamma=0$ it is easily seen from the expression of
$\widetilde{U}^{ab}$ in Eq.~(\ref{utilde}) (or equivalently from
Eq.~(\ref{Kinterm2})) that each matrix $U_{{\bm x},\mu}$ can be
multiplied by an arbitrary $({\bm x},\mu)$-dependent element of the
gauge group center without changing the action: for $\gamma=0$ the
gauge group is in fact SU$(N_c)/{\mathbb Z}_{N_c}$.  This is
responsible for the vanishing of the average value of the plaquette,
$\langle {\rm Tr}\,\Pi_{\bm x} \rangle=0$, for $\gamma=0$. Finally,
note that for $N_c = 2$ and again $\gamma = 0$, because of the
isomorphism SU$(2)/{\mathbb Z}_{2}=$SO(3), we are dealing with a
theory with SO(3) local symmetry.

In the following we consider a simplified model, which can be formally
obtained by setting $r = - 2u$, and taking the limit $u\to\infty$.
The model has fixed-length fields and a simpler potential:
\begin{equation}
  {\rm  Tr}\,\Phi_{\bm x}^t \Phi_{\bm x}^{\phantom t} = 2 \,, \qquad 
V(\Phi) = {v\over 4} \, {\rm Tr}\,\Phi^t\Phi \Phi^t\Phi\,.  
\label{unitlength}
\end{equation}
In terms of the variables $\phi^f_{ij}$ defined in Eq.~(\ref{psidef}) we have
\begin{equation}
\sum_f {\rm Tr}\,\phi_{\bm x}^f \phi_{\bm x}^f= 1\,, \qquad V(\phi) 
= v\, \sum_{fg} ({\rm Tr}\,\phi^f\phi^g)^2 \,.
\label{conphif}
\end{equation}
Therefore, we consider the action
\begin{equation}
\label{hfixedlength}  
S = - {N_f\over 2} \sum_{{\bm x},\mu} {\rm Tr} \,\Phi_{\bm x}^t \,
\widetilde{U}_{{\bm x},\mu} \, \Phi_{{\bm x}+\hat{\mu}}^{\phantom t}
+ {v\over 4} \sum_{\bm x}  {\rm Tr}\,\Phi_{\bm x}^t\Phi_{\bm x}
\Phi_{\bm x}^t\Phi_{\bm x}
- {\gamma\over N_c} \sum_{{\bm x}}  {\rm Re} \, {\rm Tr}\,
\Pi_{\bm x}.
\end{equation}
We expect this simplified model to show all 
universal features of the models with generic values of $r$ and $u$.  

The critical properties in the zero-temperature limit can be monitored
by the correlation functions of the gauge-invariant bilinear operators
\begin{equation}
  B_{\bm x}^{fg} = {1\over 2} \sum_a \Phi_{\bm x}^{af} \Phi_{\bm x}^{ag}\,,
  \qquad
  Q_{\bm x}^{fg} = B_{\bm x}^{fg} - {1\over N_f} \delta^{fg}\,,
\label{qdef}
\end{equation}
which satisfy ${\rm Tr} \, B_{\bm x}=1$ and ${\rm Tr} \, Q_{\bm x}=0$,
due to the fixed-length constraint.
Assuming translation invariance, holding for finite-size systems with
periodic boundary conditions, we define the two-point correlation
function
\begin{equation}
G({\bm x}-{\bm y}) = \langle {\rm Tr}\, Q_{\bm x} Q_{\bm y} \rangle\,,  
\label{gxyp}
\end{equation}
the corresponding susceptibility $\chi=\sum_{\bm x} G({\bm x})$
and second-moment correlation length
\begin{eqnarray}
\xi^2 = {1\over 4 \sin^2 (\pi/L)} {\widetilde{G}({\bm 0}) -
  \widetilde{G}({\bm p}_m)\over \widetilde{G}({\bm p}_m)}\,,
\label{xidefpb}
\end{eqnarray}
where $\widetilde{G}({\bm p})=\sum_{{\bm x}} e^{i{\bm p}\cdot {\bm x}} G({\bm
x})$ is the Fourier transform of $G({\bm x})$, and ${\bm p}_m = (2\pi/L,0)$.
In addition, we consider universal renormalization-group (RG) invariant
quantities, such as the ratio
\begin{equation}
R_\xi\equiv \xi/L\,,
\label{rxidef}
\end{equation}
and the Binder parameter 
\begin{equation}
U = {\langle \mu_2^2\rangle \over \langle \mu_2 \rangle^2} \,, \qquad
\mu_2 = {1\over V^2} \sum_{{\bm x},{\bm y}} {\rm Tr}\,Q_{\bm x} Q_{\bm
  y}\,,\qquad V=L^2\,.
\label{binderdef}
\end{equation}

\section{Zero-temperature limit}
\label{varsce}

Let us now discuss the expected critical behavior. We only consider
systems with $N_f\ge 3$, in which the global symmetry is nonabelian.
In this case, we do not expect a critical behavior for finite $\beta$,
but only in the zero-temperature limit.  According to the conjecture
reported in the introduction, the critical behavior should be the same
as that of the 2D $\sigma$ models defined on the symmetric spaces with
the same global symmetry, that is the models defined
on~\cite{BHZ-80,ZJ-book} O($N_f$)/O($p$)$\otimes$O($N_f-p$) for
different values of $p$.

\subsection{Zero-temperature relevant configurations} \label{sec3.1}

As a first step, we identify the 
relevant configurations for $\beta \to \infty$, which are controlled by the 
action terms  $S_K(\Phi,U)$ and $S_V(\Phi)$. As in two dimensions there is 
no critical pure-gauge dynamics, we expect, and we will verify numerically,
that $S_G(U)$ does not play a relevant role. Although we will be interested
in systems with $N_f\ge 3$, the results for the zero-temperature configurations 
also hold for $N_f=2$.

Let us first consider the potential term $S_V(\Phi)$. For $\beta\to\infty$,
the relevant configurations are those that minimize $V(\Phi)$ defined in 
Eq.~(\ref{unitlength}). To determine the minima, we use the 
singular value decomposition that allows us to rewrite the field 
$\Phi$ as 
\begin{equation}
  \Phi^{af} = \sum_{bg} C^{ab} W^{bg} F^{gf}\,,
  \label{singdec}
\end{equation}
where $C\in {\rm O}(N_c^2-1)$ and $F\in {\rm O}(N_f)$ are orthogonal matrices,
and $W$ is an $(N^2_c-1) \times N_f$ rectangular matrix with zero nondiagonal
elements ($W_{ij} = 0$ for $i\not= j$). We set
$W_{ii} = w_i$ with
$i=1,...,q$, where 
\begin{equation}
  q={\rm Min}[N_f,N_c^2-1]\,.
  \label{qqdef}
\end{equation}
Without loss of generality, we assume that $w_i \ge 0$. 
Substitution in $V(\Phi)$ gives 
\begin{equation}
   V(\Phi) = v \sum_{i=1}^q w_i^4.
\end{equation}
If we minimize $V(\Phi)$ subject to the constraint $\hbox{Tr}\, \Phi^t \Phi =
  \sum_{i=1}^q w_i^2 = 2$, it is easy to verify that there are two
solutions that depend on the sign of $v$: 
\begin{equation}
\begin{aligned}
\hbox{(I)} \quad & w_1 = \sqrt{2}, \qquad w_i = 0 \quad\hbox{for $i \ge 2$}\,,  \\
\hbox{(II)} \quad & w_1 = \ldots =  w_q = (2/q)^{1/2}\, . 
\end{aligned}
\label{soluzioni}
\end{equation}
Solution (I) is the relevant one for $v < 0$, while 
solution (II) is the relevant one for $v > 0$. It is interesting to observe
that this result also holds for the general potential (\ref{potential}), as
long as $r < 0$. For $r> 0$, the minimum of the potential corresponds to 
$w_1 = \ldots w_q = 0$: no critical behavior is expected in this case. 

For solutions of type (I), we can rewrite the field as 
\begin{equation}
  \Phi^{af} = \sqrt{2} s^a z^f, 
\label{Fieldbetainf-1}
\end{equation}
where $s$ and $z$ are unit real vectors of dimension 
$N^2_c - 1$ and $N_f$, respectively. For solutions of type (II), we have
instead  
\begin{equation}
\Phi^{af} = 
\sqrt{2\over q} \sum_{k=1}^q C^{ak} F^{kf}.
\label{Phi-vgt0}
\end{equation}
This expression can be simplified, parametrizing $\Phi$ 
in terms of a single orthogonal matrix. We should distinguish two 
cases. If $N_f\ge N^2_c - 1=q$, 
let us define an $N_f$-dimensional orthogonal 
matrix $\widehat{C} = C\oplus I_{N_f-q}$, 
where $I_{p}$ is the $p$-dimensional identity matrix.  We can rewrite 
Eq.~(\ref{Phi-vgt0}) as 
\begin{equation}
\Phi^{af} = 
\sqrt{2\over q} \sum_{g=1}^{N_f} \widehat{C}^{ag} F^{gf} .
\end{equation}
Since $\widehat{C}$ is an orthogonal matrix, we can express $\Phi$ in
terms of a single orthogonal matrix $F' = \widehat{C} F$, i.e., we can
set $C = I$ in Eq.~(\ref{Phi-vgt0}). Of course, because of gauge
invariance, see Eq.~(\ref{transformPhi}), $F$ is not uniquely defined
and it is, more properly, an element of O($N_f$)/SU($N_c)_{\rm adj}$
[SU($N_c)_{\rm adj}$ is the group of block-diagonal matrices
  $\widetilde{V}\oplus I_{N_f-q}$, where $\widetilde{V}$ belongs to
  the adjoint representation of SU($N_c$)].  For $N_c=2$ the quotient
becomes SO($N_f$)/SO(3).  If $N_f \le N^2_c - 1$, we can repeat the
same argument to prove that one can set $F = I$ and $\Phi^{af} =
C^{af}$, without loss of generality. Note that, for $N_c = 2$, we can
use the gauge transformations to further simplify the field. Indeed,
the matrix $\widetilde{V}$ appearing in Eq.~(\ref{transformPhi}) is a
generic orthogonal matrix. Thus, choosing $\widetilde{V} = C^t$, we
obtain $\Phi^{af} = \delta^{af}$: the minimum-potential field
configuration is completely determined.

In the previous calculation we have assumed that the relevant
scalar-field configurations in the large-$\beta$ limit are only
determined by the potential term $S_V(\Phi)$. In Appendix \ref{AppA},
we discuss the role of the kinetic term $S_K(\Phi,U)$ and show that
this quantity is not relevant for the determination of the
low-temperature behavior of the scalar field for $v\not=0$. The
kinetic term is only relevant for $v=0$.  In this case, we can show
that, for $N_c=2$, the model with $v=0$ behaves as for $v < 0$ (see
App.~\ref{AppA}): the relevant configurations correspond to solution
(I) reported above.  We do not have exact results for $N_c >
2$. However, the numerical results we will present below indicate that
also for $N_c > 2$, the relevant configurations for $v = 0$ are those
of type (I).

To distinguish the nature of the
zero-temperature configurations, one can use the 
order parameter $B_{\bm x}$ defined in Eq.~(\ref{qdef}). If the field 
is parametrized as in Eq.~(\ref{singdec}), we have
\begin{equation}
\hbox{Tr}\, B^2 =\, {1\over 4} \sum_{i=1}^q w_i^4,
\end{equation}
so that 
\begin{eqnarray}
\hbox{(I)} && \qquad \hbox{Tr}\, B^2 = 1, \nonumber \\
\hbox{(II)}&& \qquad \hbox{Tr}\, B^2 = {1\over q},
\label{B2-predictions}
\end{eqnarray}
for solutions of type (I) and (II), respectively [see Eq.~(\ref{soluzioni})]. 

It is interesting to note that in this discussion the gauge group does
not play any role: the only relevant quantity is the dimension of the
gauge representation. In particular, one would obtain exactly the same
results for the minimum configuration and the behavior of the order
parameter $Q$ for a gauge theory in which the fields transform under
the fundamental representation of the O($N_c^2-1$) group.

In the previous discussion, we focused on the minimum
configurations of the scalar fields.  We wish now to discuss the large-$\beta$
behavior of the gauge fields.  If we minimize the kinetic term (\ref{Kinterm}),
we obtain 
\begin{equation}
\Phi_x = \widetilde{U}_{x,\mu} \Phi_{x+\hat{\mu}}.
\end{equation}
Repeated applications of this relation along a plaquette give
\begin{equation}
\Phi_x = \widetilde{\Pi}_x \Phi_x \qquad 
\widetilde{\Pi}_{\bm x}=
      \widetilde{U}_{{\bm x},1} \,\widetilde{U}_{{\bm x}+\hat{1},2} 
     \,\widetilde{U}_{{\bm x}+\hat{2},1}^t \,\widetilde{U}_{{\bm x},2}^t .
\end{equation}
For minimum configurations of type (I), using Eq.~(\ref{Fieldbetainf-1}), we 
have 
\begin{equation}
s^{a} = \sum_b \widetilde{\Pi}^{ab} s^b ,
\end{equation}
i.e., $\widetilde{\Pi}_x$ has necessarily a unit eigenvalue. 
A detailed analysis shows that 
$\widetilde{\Pi}_x$ can be written as $\exp(i\sum_a \alpha^a \widetilde{T}^a)$,
where $\widetilde{T}^a$ are the generators in the adjoint representation 
of a smaller subgroup isomorphic to $U(1)\oplus U(N_c-2)$. Thus, for 
$\beta\to\infty$ there is still a residual dynamics of the gauge fields, 
i.e., we end up with a $U(1)\oplus U(N_c-2)$ pure gauge model with 
Hamiltonian $H_G(U)$. In two dimensions, however, this dynamics is unable to 
give rise to a critical behavior. 

Let us now consider the case in which the relevant configurations are
those of type (II), see Eq.~(\ref{soluzioni}). In this case,
$\widetilde{\Pi}$ has $q$ unit eigenvalues, which further reduce the
dynamics of the gauge fields.  In particular, for $N_f \ge N_c^2 - 1$,
$\widetilde{\Pi}_x = 1$.  Note, however, that this still leaves open
the possibility of a nontrivial dynamics for the fields $U_x$. Indeed,
the condition $\widetilde{\Pi}_x = 1$ implies that $\Pi_x$ belongs to
the center ${\mathbb Z}_{N_c}$ of the group, so that, in the limit
$\beta \to \infty$, we end up with a ${\mathbb Z}_{N_c}$ pure gauge
theory. Again, as we are in two dimensions, this gauge model is not
expected to become critical and therefore it should not be relevant
for the critical dynamics of the model.

\subsection{Effective models for the low-temperature behavior} \label{sec3.2}

Let us now analyze the effective behavior in the zero-temperature
limit.  We first assume that $v$ is negative or vanishes, so that the
relevant minimum configurations are those of type (I).  Then, we
assume that, for $\beta \to \infty$, the relevant fluctuations are
those that locally satisfy the minimum potential conditions, i.e.,
that we can parametrize the field as in Eq.~(\ref{Fieldbetainf-1})
with site dependent vectors $z_{\bm x}$ and $s_{\bm x}$.  The field
(\ref{Fieldbetainf-1}) satisfies the minimum condition
exactly. Fluctuations are possible as we do not assume translation
invariance, so that $z_{\bm x}$ and $s_{\bm x}$ are site
dependent. For this type of field configurations the kinetic term
becomes
\begin{equation}
S_K = - J N_f \sum_{{\bm x},\mu} j_{{\bm x},\mu} 
    {\bm z}_{\bm x}\cdot {\bm z}_{{\bm x}+\hat{\mu}} , \qquad
j_{{\bm x},\mu} = \sum_{ab} s_{\bm x}^a 
{\widetilde U}^{ab}_{{\bm x},\mu} s_{{\bm x}+\hat{\mu} }^b .
\end{equation}
It is trivial to see that the coupling $j_{{\bm x},\mu}$ satisfies
$|j_{{\bm x},\mu}| \le 1$.  For large values of $\beta$, $S_K$ should
be minimized, which requires either ${\bm z}_{\bm x} = {\bm z}_{{\bm
    x}+\hat{\mu}}$ and $j_{{\bm x},\mu}=1$ or ${\bm z}_{\bm x} = -{\bm
  z}_{{\bm x}+\hat{\mu}}$ and $j_{{\bm x},\mu}=-1$.  As these two
possibilities occur with the same probability, the effective model for
the fluctuations is a gauge RP$^{N_f-1}$ model, in which $j_{{\bm
    x},\mu}$ is a gauge field that takes the values $\pm 1$ with equal
probability.

At a more intuitive level, the correspondence between the critical
behavior of the gauge model and of the RP$^{N_f-1}$ model can be
established by noting that the order parameter $Q$, or equivalently
$B$, defined in Eq.~(\ref{qdef}), can be written as
\begin{equation}
B_{\bm x}^{fg} = z_{\bm x}^f z_{\bm x}^g = P^{fg}_{\bm x} ,
\end{equation}
which shows that $B_{\bm x}$ is a local projector $P^{fg}$ 
onto a one-dimensional space. If 
we assume that the dynamics in the gauge model is completely determined 
by the fluctuations of the order parameter $B_{\bm x}$, we immediately identify 
the effective scalar model as the RP$^{N_f-1}$ model. Indeed, 
the standard nearest-neighbor
RP$^{N-1}$ action is obtained by taking the simplest action for a
local projector $P^{fg}_{\bm x}$:
\begin{eqnarray}
  S_{\rm RP} = - J \sum_{{\bm x},\mu} \hbox{Tr}\,
      P_{\bm x} P_{{\bm x}+\hat{\mu}}\,,
  \qquad P_{\bm x}^{fg} = \varphi_{\bm x}^f \varphi_{\bm x}^g\,,
\label{srp}
\end{eqnarray}
where $\varphi^a_{\bm x}$ is a unit vector and $P_{\bm x}^{ab} =
\varphi^a_{\bm x} \varphi^b_{\bm x} $ is a local projector onto a
one-dimensional space, i.e., it satisfies $P_{\bm x} = P_{\bm x}^2$
and ${\rm Tr} P_{\bm x} =1$.  The zero-temperature critical behavior
of 2D RP$^{N-1}$ models is still debated; see, e.g.,
Refs.~\cite{CEPS-93, CPS-94, Hasenbusch-96, NWS-96, CHHR-98, DDL-20,
  BFPV-20}.  Although 2D RP$^{N-1}$ and O($N$) $\sigma$ models have
the same perturbative behavior~\cite{Hasenbusch-96}, there is
numerical evidence that their nonperturbative behavior differs.  This
is due to topological ${\mathbb Z}_2$ defects that are present in the
RP$^{N-1}$ model, which are apparently relevant perturbations of the
zero-temperature 2D O($N$) fixed point, leading to a different
universal asymptotic behavior in the nonperturbative
regime~\cite{BFPV-20}.

Let us now assume $v > 0$. The relevant solutions are those of type
(II). In this case, we must distinguish two cases.  If $N_f \le N_c^2
- 1$ we find
\begin{equation}
  B = {1\over 2} F^t W^t W F = {1\over N_f} I_{N_f} ,
  \label{btri}
\end{equation}
where $I_{N_f}$ is the $N_f$-dimensional unit matrix. Correspondingly,
the order parameter $Q$ vanishes in the limit $\beta\to
\infty$. Therefore, correlations of $Q_x$, and also the Binder
parameter, depend on the fluctuations of the field $\Phi^{af}_{\bm x}$
around the minimum configurations.  We do not have predictions for
their behavior. However, we will show numerically below that these
fluctuations do not show long-range correlations. Indeed, for $T\to
0$, the Binder parameter takes the high-temperature value appropriate
for disordered configurations:
\begin{equation}
  \lim_{\beta\to 0} U = 1 + {4\over (N_f-1)(N_f+2)}\, .
  \label{Ubeta0}
\end{equation} 
Note that this result is consistent with what we assumed for $v < 0$:
fluctuations around the minimum configurations are irrelevant and the critical 
behavior is only due to the fluctuations of the 
fields that locally minimize the 
quartic potential. In the case we are discussing now, once the fields minimize
the quartic potential, the order parameter $Q$ is fixed---it vanishes---and 
therefore no long-range fluctuations of $Q_{\bm x}$ are possible. 

Let us finally suppose that $N_f > N_c^2 - 1$. In this case, the order
parameter is nontrivial and the system orders. To identify the
effective model, note that $\widetilde{\Pi} = 1$, so that
$\widetilde{U}_{{\bm x},\mu} = \widetilde{V}_{\bm x}
\widetilde{V}_{{\bm x}+\hat{\mu}}$.  As in the discussion for $v < 0$,
we assume that the fields locally minimize the potential, so that
$\Phi^{af}_{\bm x}$ can be parametrized as in Eq.~(\ref{Phi-vgt0})
with $C = I$ and a site-dependent orthogonal matrix $F_{\bm x}$.
Substituting this parametrization in the kinetic term of the action we
obtain
\begin{equation}
S_K = - {N_f\over q} \sum_{{\bm x}\mu} \hbox{Tr}\, 
  (F^t_{\bm x}  \widehat{V}_{\bm x} Y_{N_f}^q 
   \widehat{V}_{{\bm x}+\hat{\mu}}^t F_{{\bm x}+\hat{\mu}} ) ,
\end{equation}
where\footnote{We indicate with $A\oplus B$ a block-diagonal matrix,
where $A$ and $B$ are square matrices of dimension $q$ and $N_f-q$,
respectively.} $Y_{N_f}^q = I_{q} \oplus 0$ is an $N_f\times N_f$
diagonal matrix in which the first $q$ elements are 1 and the other
$(N_f-q)$ elements are 0 and $\widehat{V} = \widetilde{V} \oplus
I_{N_f-q}$.  Note that action is invariant under SU($N_c)_{\rm
  adj}\otimes$O($N_f-q$) transformations defined by $F \to W_F F$,
$\widehat{V} \to W_V \widehat{V}$, where $W_F = W_1 \oplus W_2$, $W_V
= W_1 \oplus I$, with $W_1 \in \hbox{SU($N_c)_{\rm adj}$}$ and $W_2
\in \hbox{O($N_f-q$)}$. As we already mentioned in Sec.~\ref{sec3.1},
we expect the same critical behavior if we consider O($q$) gauge
fields, leading to an effective enlargement of the symmetry to
O($q$)$\otimes$O($N_f-q$).  The resulting effective model is therefore
a lattice $\sigma$ model defined on the symmetric space
O($N_f$)/O($q$)$\otimes$O($N_f-q$) \cite{BHZ-80,ZJ-book}.  For $N_f =
q + 1$ the symmetric space is isomorphic to the sphere in $N_f$
dimensions, and thus the effective model is simply the
O($N_f$)-invariant vector $\sigma$ model with Hamiltonian
\begin{eqnarray}
  S_{\rm O(N)} = - J \sum_{{\bm x},\mu} 
  \varphi_{\bm x} \cdot \varphi_{{\bm x}+\hat{\mu}} \,,
  \qquad \varphi_{\bm x}\cdot \varphi_{\bm x}=1\, .
\label{srpon}
\end{eqnarray}

\subsection{Numerical results}

\begin{table}
  \centering
  \begin{tabular}{cclll}
\hline\hline
    $N_c$ & $\quad N_f \quad$ & $\;\langle {\rm Tr} \, B^2
    \rangle_{v=0}$& $\;\langle {\rm Tr} \, B^2 \rangle_{v=1}$ &
    $\;\;U_{v=1}$ \\ \hline
    2 & 2 &0.9998(4)&0.49981(13) & 2.000(2) \\
    2 & 3 &0.9999(5)&0.3331(2) & 1.4014(7)\\
    2 & 4 &1.0000(4)&0.3332(2)& 0.9997(2) \\
    2 & 5 &         &0.3333(5)& 0.9999(2) \\
    2 & 6 &         &0.3335(4)& 1.0000(1) \\\hline
    3 & 2 &1.0006(12)&0.4999(5) & 2.012(5) \\
    3 & 3 &0.9999(12)&0.3327(12)& 1.401(2) \\
    3 & 4 &1.0002(14)&0.251(2)& \\\hline
    4 & 2 &0.99(2)&0.500(3) & 1.991(6) \\
    4 & 3 &1.000(2)&0.330(5)& 1.407(3) \\
    4 & 4 &0.999(3)&0.236(8)& \\     
    \hline \hline
  \end{tabular}
  \caption{Results for 
$\langle {\rm Tr} \, B^2_{\bm x} \rangle$ and the Binder parameter $U$ in the 
large-$\beta$ limit. We consider a square lattice of
size $L=4$, two values of $v$, $v=0$ and $v=1$, and $\gamma=0$. 
The values of $U_{v=1}$ that differ from 1 are consistent 
with Eq.~(\ref{Ubeta0}), which predicts $U=2$ and 7/5 for $N_f=2$ and 3, 
respectively.}
\label{v01minima}
\end{table} 

To verify the above-reported predictions, we have performed 
MC simulations
(see App.~\ref{MCsim} for details) for large values of $\beta$, $\gamma=0$
(the gauge action is not expected to play an important role),
and relatively small systems.  The extrapolations of the results provide
information on the nature of the relevant low-temperature configurations.
In Table~\ref{v01minima} we report the large-$\beta$ extrapolations of 
$\langle {\rm Tr}\,B_{\bm x}^2\rangle$ for 
$v=0$ and $v=1$.  The results should be compared with the prediction
(\ref{B2-predictions}). For $v=0$, the average is always consistent with 1,
confirming that the relevant configurations correspond to solution (I).
Apparently, for any $N_c$, for $v=0$ the model behaves as for $v <
0$, a result that we have only proved for $N_c=2$. For $v=1$, results are 
instead consistent with $1/q$, confirming that the relevant configurations are 
those of type (II). Note that this result also applies when $N_f=2$. 
In this case, the flavor symmetry is abelian and therefore a BKT
finite-temperature transition is possible.

We also computed the Binder parameter $U$. For $v=0$, it always converges to 
1 as $\beta \to \infty$ (data not shown), indicating that long-range
correlations set in in the limit. Results for $v=1$ are reported in 
Table~\ref{v01minima}. For $N_f \le N^2_c - 1$ (all results we report for 
$N_c = 3,4$ satisfy this condition) we observe that $U$ is always approximately 
equal to the 
high-temperature value (\ref{Ubeta0}), $U = 2$ and 7/5 for $N_f = 2$ and 
3, respectively. This indicates the absence of long-range correlations. 
In the opposite case $N_f > N^2_c - 1$, we find instead $U=1$, consistent 
with the presence of critical fluctuations.

\subsection{Summary}

To conclude the Section, let us summarize the expected low-temperature 
behavior of the model for $N_f \ge 3$:

\noindent
i) For $v \le 0$, the model has a zero-temperature 
critical (continuum) limit independent of $N_c$, analogous to that of the
RP$^{N_f-1}$ $\sigma$ model.

\noindent
ii) For $v > 0$ and $N_f \le N^2_c - 1$, $Q$ correlations are always
short-ranged, even in the limit $T\to 0$.

\noindent
iii) For $v > 0$ and $N_f > N^2_c - 1$ the model has a zero-temperature 
critical (continuum) limit that depends both on $N_f$ and $N_c$. It is the same
as that of the $\sigma$ model defined on the symmetric space 
\cite{BHZ-80,ZJ-book} O($N_f$)/O($q$)$\otimes$O($N_f-q$): correlations of the 
order parameter $Q$ have the same critical behavior, i.e., continuum
limit, in the two models.  For 
$N_f = N^2_c$, we obtain the same behavior as that of the O($N_f$) vector
$\sigma$ model.

Note that these effective behaviors have been obtained by making very simple
assumptions. Essentially, we have assumed that the relevant 
configurations correspond to scalar fields $\{\Phi^{\rm min} \}$ that locally,
---i.e., at each site---minimize the potential $S_V(\Phi)$. 
Gauge fields are only relevant for the identification of the 
dynamic degrees of freedom and for restricting the focus on gauge-invariant
observables. As a consequence, we expect that the model presented here has 
the same continuum limits of the model with fields that transform 
under the fundamental representation of the O($N_c^2-1$) gauge group.

\section{Numerical results}
\label{numres}

In this section we present some numerical results that confirm the
predictions of Sec.~\ref{varsce}. Results for $v = 0$ and $N_c = 2$
have been already discussed in Refs.~\cite{BFPV-20-ong} ($N_f = 3,4$) and 
\cite{BFPV-21} (where $N_f=2$, so that a finite-temperature BKT transition
was observed). Indeed, for $N_c=2$ the model can be rewritten as 
an SO(3) gauge theory with fields in the fundamental representation of the 
gauge group. The results presented in Ref.~\cite{BFPV-20-ong} 
are in full agreement
with the analysis presented here. In the following, we will present results
for $N_c = 3$, $v=0$  and for $N_c=2,3$, $v > 0$. 
They fully confirm the conclusions of Sec.~\ref{varsce}.

\subsection{Finite-size scaling analysis}
\label{FSSstrat}

To identify the nature of the zero-temperature critical behavior, we
perform a FSS analysis of the MC data.  We follow the
strategy already employed in
Refs.~\cite{BPV-19-ah2,BPV-20-qcd2,BFPV-20-ong,BFPV-21}. In the FSS
limit $L\to\infty$ at $L/\xi$ fixed, the RG invariant
quantities $R_\xi$ and $U$, defined in Eqs.~(\ref{rxidef}) and
(\ref{binderdef}), respectively, are expected to scale as 
\begin{equation}
U(\beta,L) \approx F_U(R_\xi)\,,
\label{r12sca}
\end{equation}
where $F_U(x)$ is a universal function that completely characterizes the
universality class of the transition.  Because of the universality of relation
(\ref{r12sca}), one may use plots of $U$ versus $R_\xi$ to identify the models
that have the same universal behavior.  If the estimates of $U$ for two
different systems approach the same curve as $L\to\infty$ when plotted versus
$R_\xi$, the transitions in the two models belong to the same universality
class. We will apply this approach below to several different models.

\subsection{Universal RP$^{N_f-1}$ behavior for $v=0$}
\label{vle0}

\begin{figure}
  \centering
  \includegraphics*[width=0.75\textwidth]{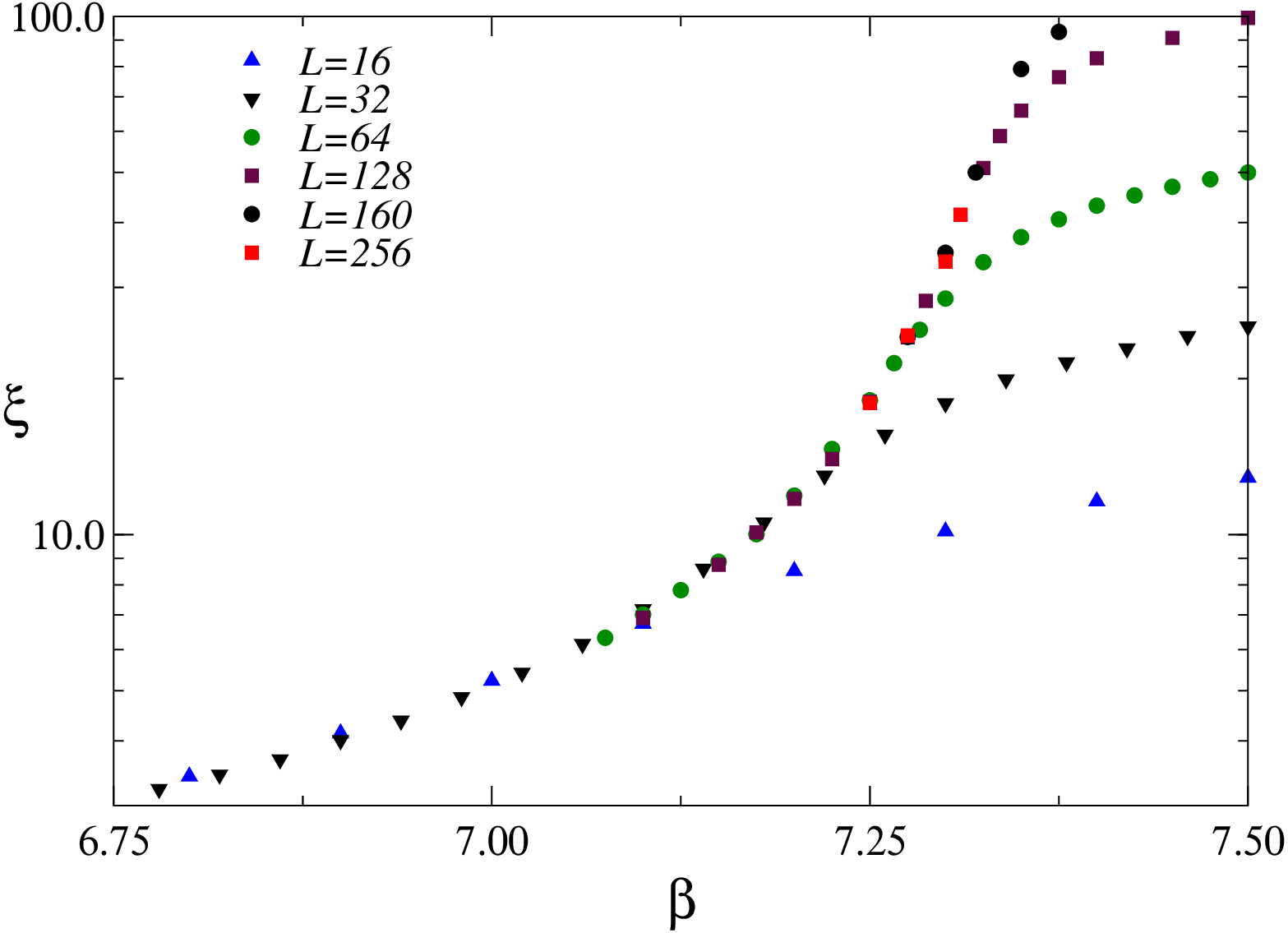}
  \caption{Estimates of $\xi$ versus $\beta$ for $N_c = 3$, $N_f = 3$, 
  $v = 0$, and $\gamma = 0$.  Results for several values of $L$ up to 256.
}
\label{nc3nf3xiv0}
\end{figure}

\begin{figure}
 \centering
 \includegraphics*[width=0.75\textwidth]{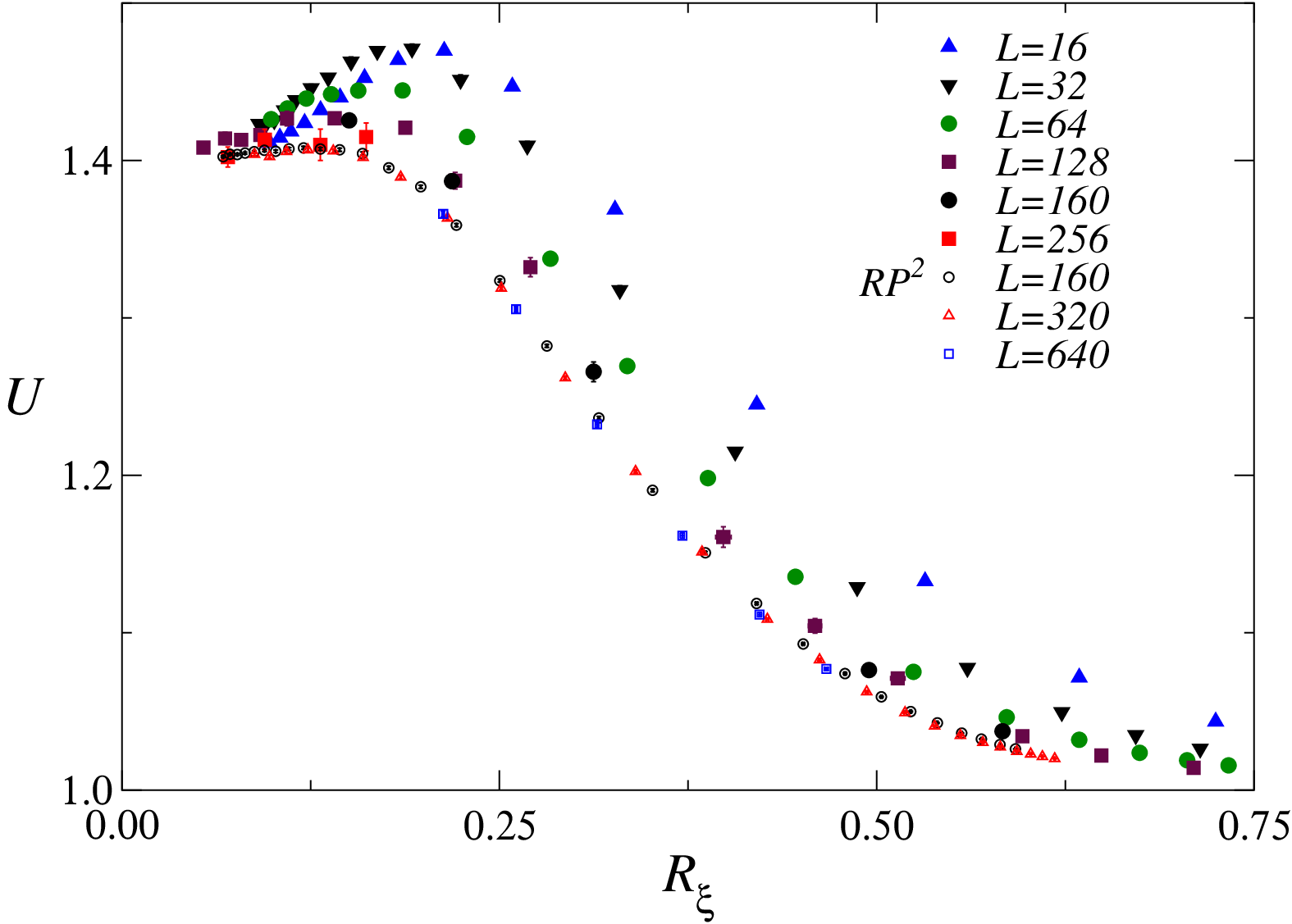}
 \caption{Plot of $U$ versus $R_{\xi}$ for $N_c = 3$, $N_f = 3$,
  $v = 0$, and $\gamma = 0$.
  Data are compared with analogous data on large lattices computed in 
  the RP$^2$ model~\cite{BFPV-20}.}
\label{nc3nf3urxiv0}
\end{figure}

\begin{figure}
\centering
\includegraphics*[width=0.75\textwidth]{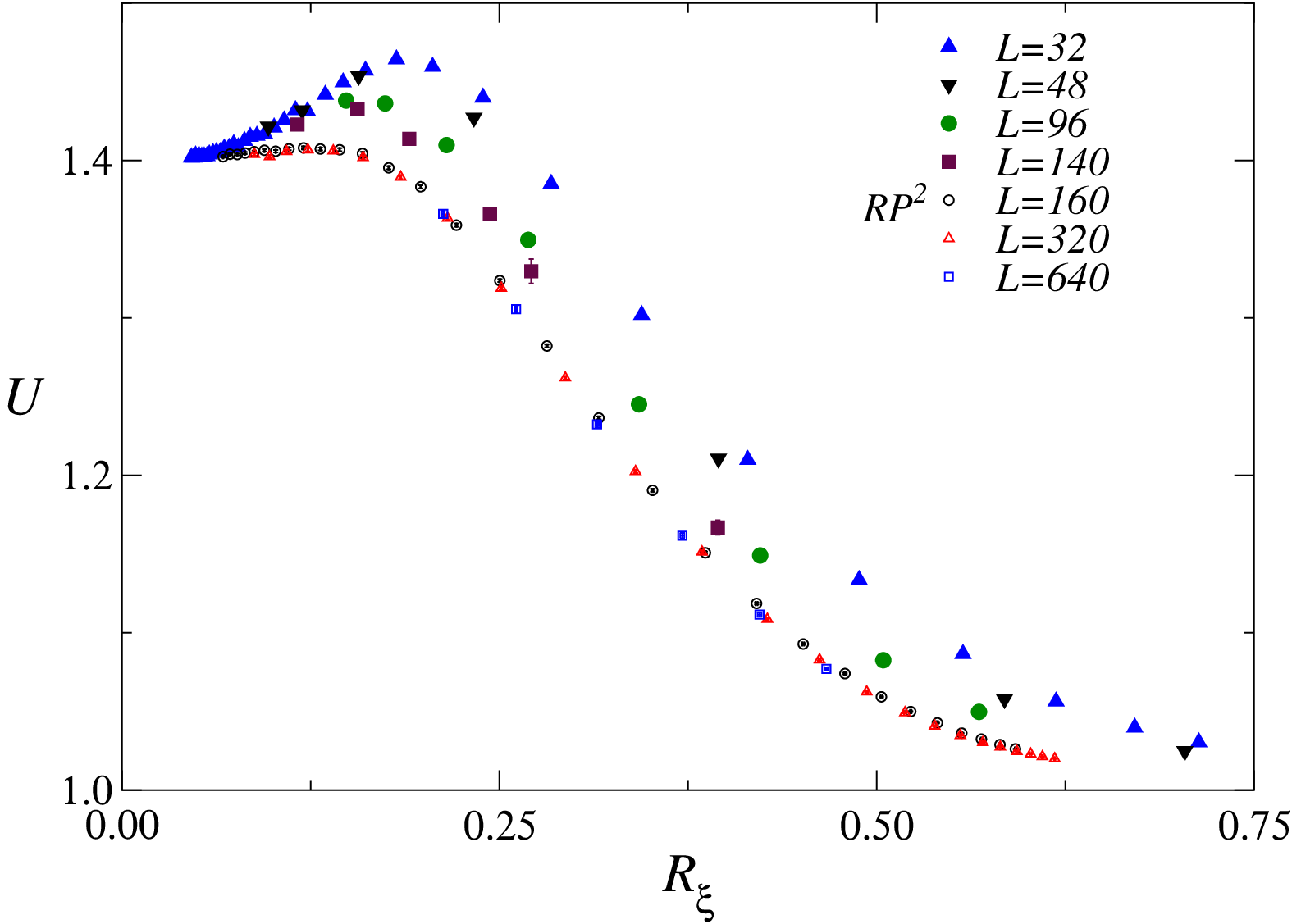}
\caption{Plot of $U$ versus $R_{\xi}$ for $N_c = 3$, $N_f = 3$,
  $v = 0$, and $\gamma = 1$.
  Data are compared with analogous data on large lattices computed in 
  the RP$^2$ model~\cite{BFPV-20}. }  
\label{nc3nf3urxiv0ga1}
\end{figure}

Let us start by considering the model for $v=0$. The model should have an
asymptotic
zero-temperature behavior analogous to that of the RP$^{N_f-1}$ model.  
In Ref.~\cite{BFPV-20-ong}, this was verified for $N_c=2$. We consider here
$N_c=3$ and $N_f=3$.
Numerical results for $\gamma=0$ are
reported in Figs.~\ref{nc3nf3xiv0} and \ref{nc3nf3urxiv0}.  The 
correlation length $\xi$ 
rapidly increases with increasing $\beta$, see Fig.~~\ref{nc3nf3xiv0},
consistently with an
asymptotic exponential behavior $\xi\sim \exp(c\beta)$. The plot of
$U$ versus $R_\xi$ reported in Fig.~\ref{nc3nf3urxiv0} shows that the data
approach the universal curve of RP$^2$ model with increasing $L$, as expected.
Scaling corrections are visible in Fig.~\ref{nc3nf3urxiv0}, but they
rapidly decay to zero (apparently as $1/L$, in the range of values of $L$ 
we consider). As we mentioned in Sec.~\ref{varsce}, the inclusion of the 
plaquette action should not change the asymptotic behavior. To verify this
point, we performed simulations with $\gamma=1$. Results are shown in 
Fig.~\ref{nc3nf3urxiv0ga1}. Also in this case, the estimates of $U$ versus 
$R_\xi$ converge towards the RP$^2$ universal curve.

\subsection{Behavior for $v>0$ and $N_f\le N_c^2-1$}
\label{vg1nfs}

\begin{figure}
  \centering
  \includegraphics*[width=0.75\textwidth]{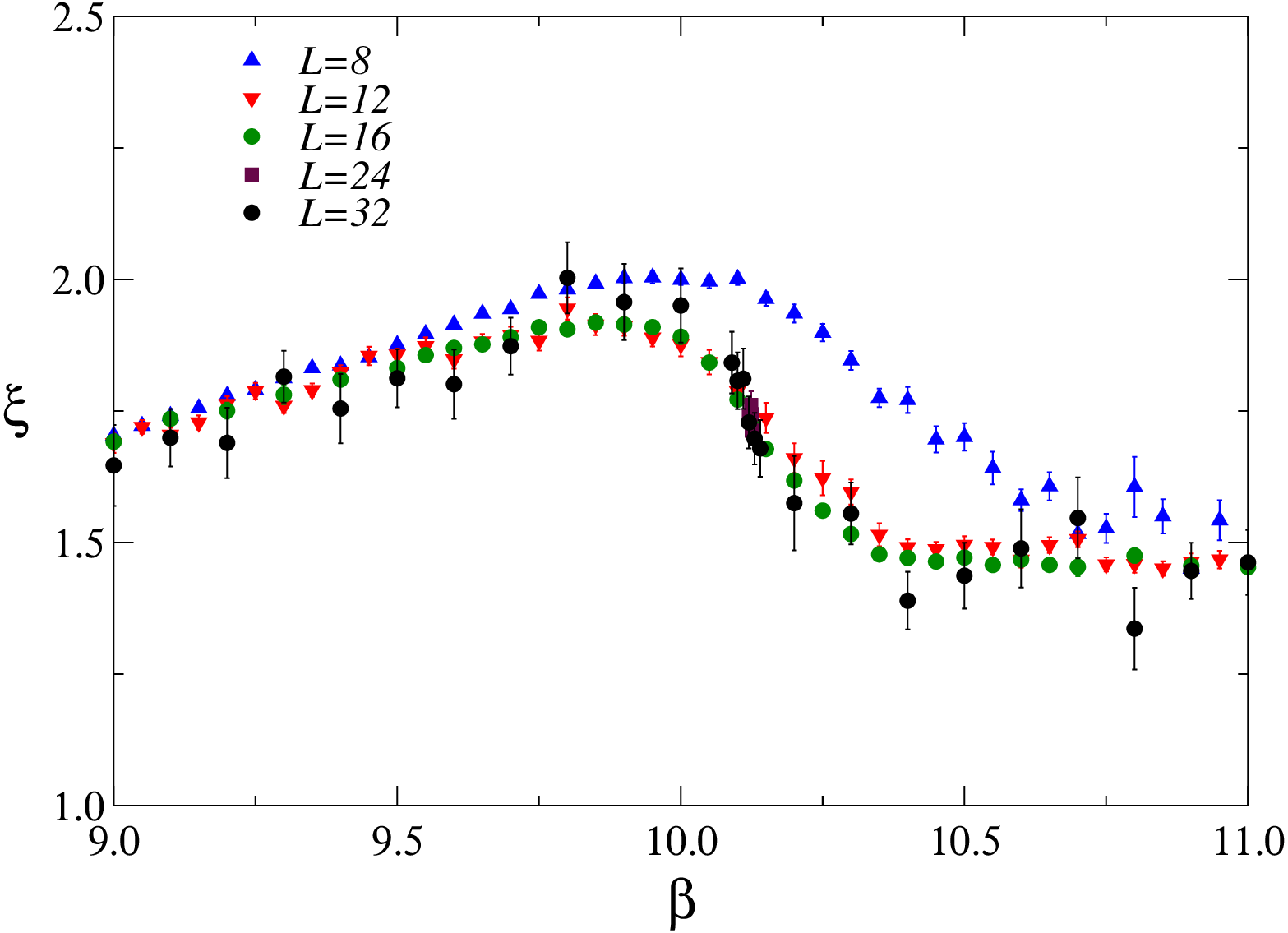}
\caption{Plot of $\xi$ versus $\beta$ for $N_c=3$, $N_f=3$, 
$\gamma=0$, and $v=1$.}
\label{fignc3nf3v1}
\end{figure}

As discussed in Sec.~\ref{varsce}, for $v>0$  and $N_f\le N_c^2-1$, 
we do not expect the correlations of the order parameter $Q$ to become 
critical for $\beta \to \infty$. Therefore, the correlation length 
should be bounded in the limit. To verify this prediction, we performed
simulations for $N_f=3$, $N_c=3$, $\gamma = 0$, and $v = 1$.
In Fig.~\ref{fignc3nf3v1} we report the correlation length as 
a function of $\beta$. It does not increase with $\beta$ and apparently 
$\xi \approx 1.5$ in the asymptotic regime. 
The data confirm that the modes associated with the scalar fields 
are disordered.  This
is also confirmed by the data of $U$, which is close to the high-temperature 
value 7/5, see Eq.~(\ref{Ubeta0}).

\subsection{Behavior for $v>0$ and $N_f>N_c^2-1$}

\label{vg1nfl}

We finally consider the model for $v>0$ and 
$N_f>N_c^2-1$.  The analysis reported in Sec.~\ref{varsce} predicts that the 
the asymptotic zero-temperature behavior is the same as that of the 
$\sigma$ model defined on the 
symmetric space O($N_f$)/O($q$)$\otimes$O($N_f-q$) with
$q=N_c^2-1$. For $N_f=N_c^2$ this is equivalent to the standard 
O($N_f$) $\sigma$ model. 

\begin{figure}
  \centering
  \includegraphics*[width=0.75\textwidth]{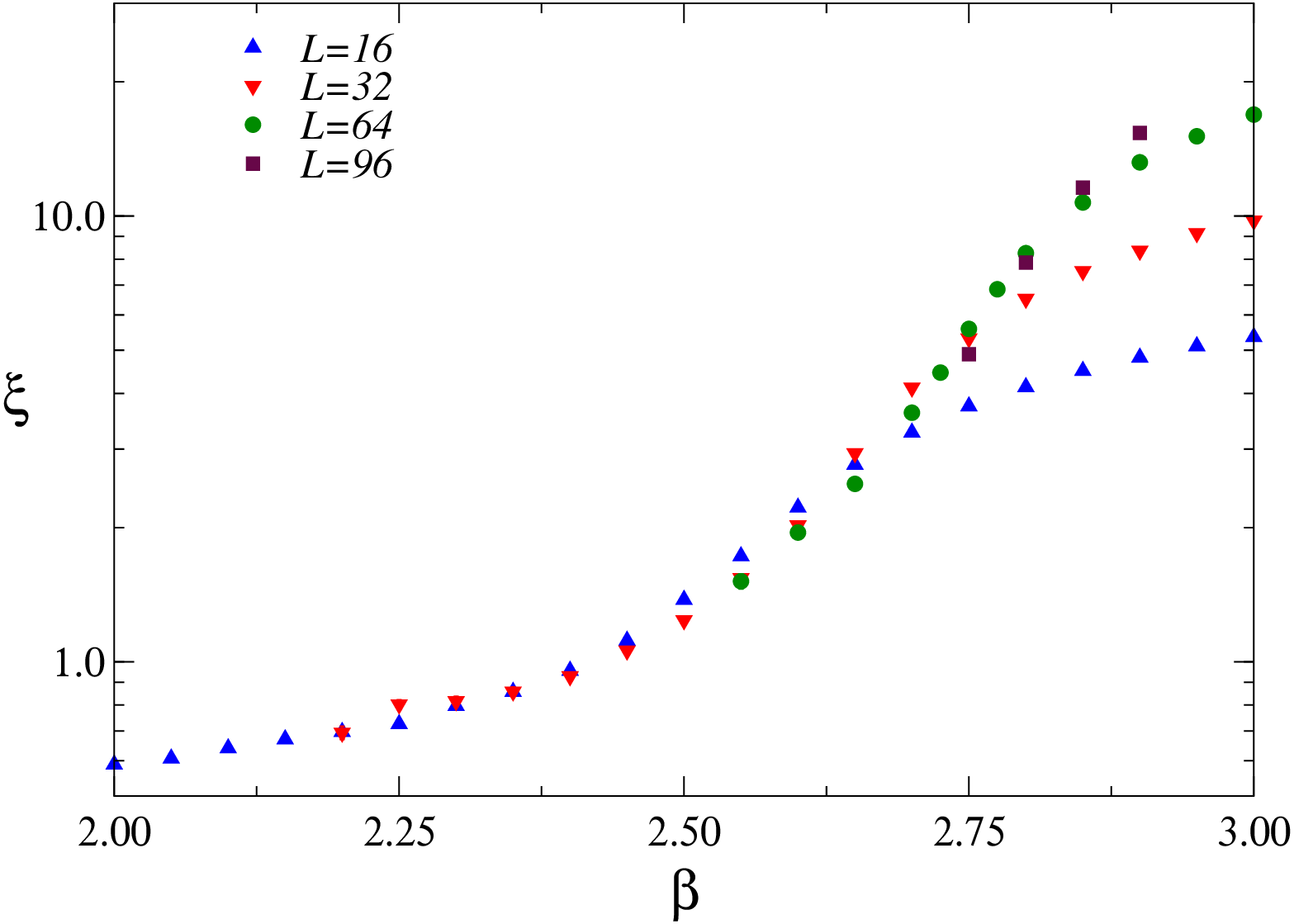}
  \caption{Correlation length $\xi$ versus $\beta$ for $N_c = 2$, $N_f=4$,
  $\gamma=0$, and $v=10$.}
\label{nf4nc2v10xi}
\end{figure}

\begin{figure}
  \centering
  \includegraphics*[width=0.75\textwidth]{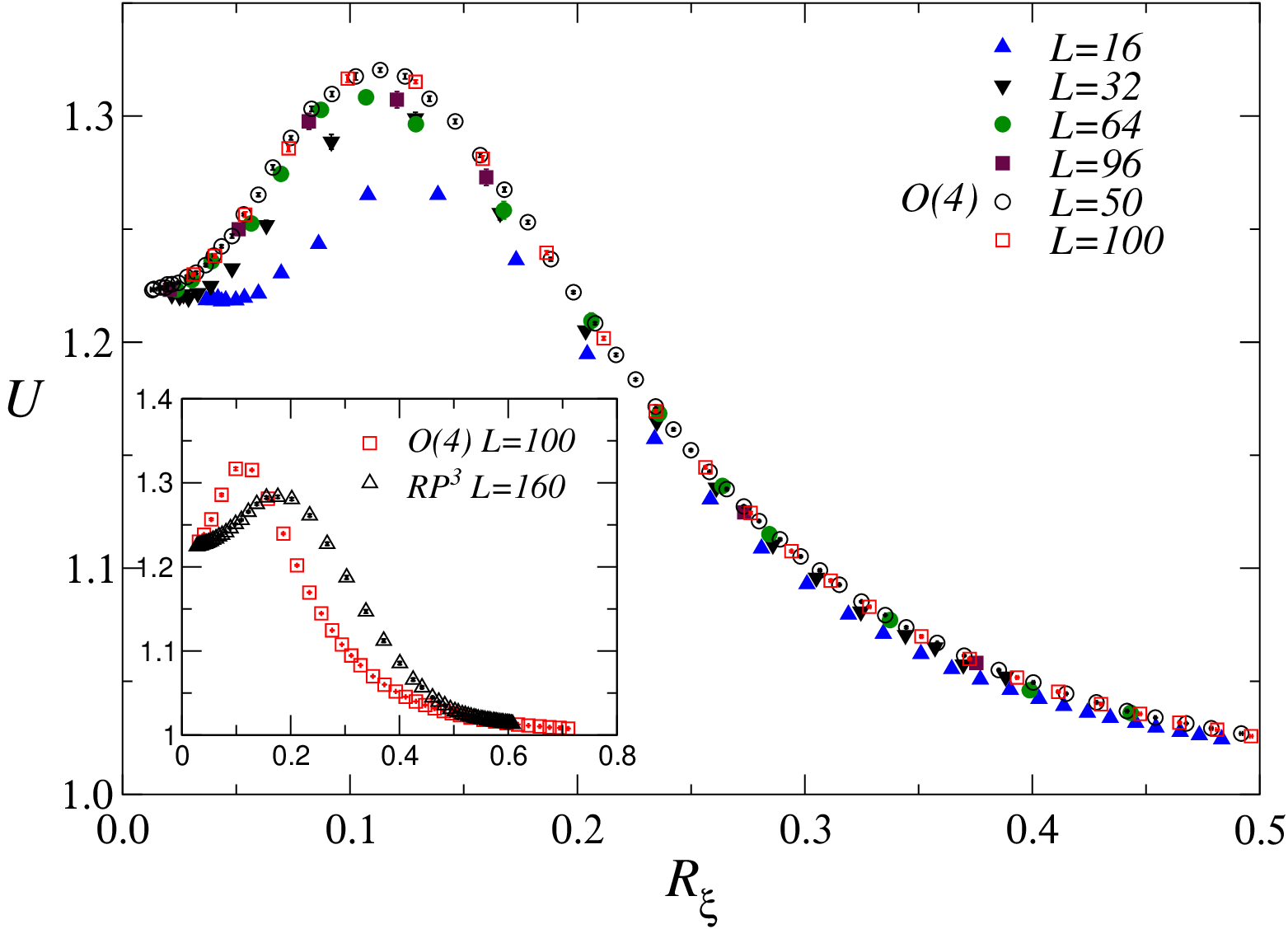}
  \caption{Plot of $U$ versus $R_\xi$ for $N_c = 2$, $N_f=4$, 
  $\gamma=0$, and $v=10$. Data 
  are compared with analogous results for the spin-2 operator 
  computed in the lattice  O(4) $\sigma$ model  with action
  (\ref{srpon}). For comparison, we also report data for the 2D
  RP$^3$ model~\cite{BFPV-20}.
}
\label{nf4nc2Urxiv10}
\end{figure}

To verify these predictions, we performed simulations for $N_f=4$, $N_c=2$,
$v = 10$, and $\gamma=0$.
In Fig.~\ref{nf4nc2v10xi} we show the correlation
length $\xi$ versus $\beta$. It increases with
increasing $\beta$, exponentially in the region in which $\xi \ll L$.
To identify the universality class, we again consider
$U$ versus $R_\xi$. The data are reported in 
Fig.~\ref{nf4nc2Urxiv10}. They appear to approach the corresponding ones
computed in the O(4) vector $\sigma$-model. Note that in the O(4) model one
should consider the same operator as in the gauge theory. Thus, the correlation
length and the Binder parameter were computed considering 
the correlation functions of the 
spin-2 operator $\varphi_{\bm x}^f \varphi_{\bm x}^g - \delta^{fg}/N_f$ with
periodic boundary conditions.

\begin{figure}
  \centering
  \includegraphics*[width=0.75\textwidth]{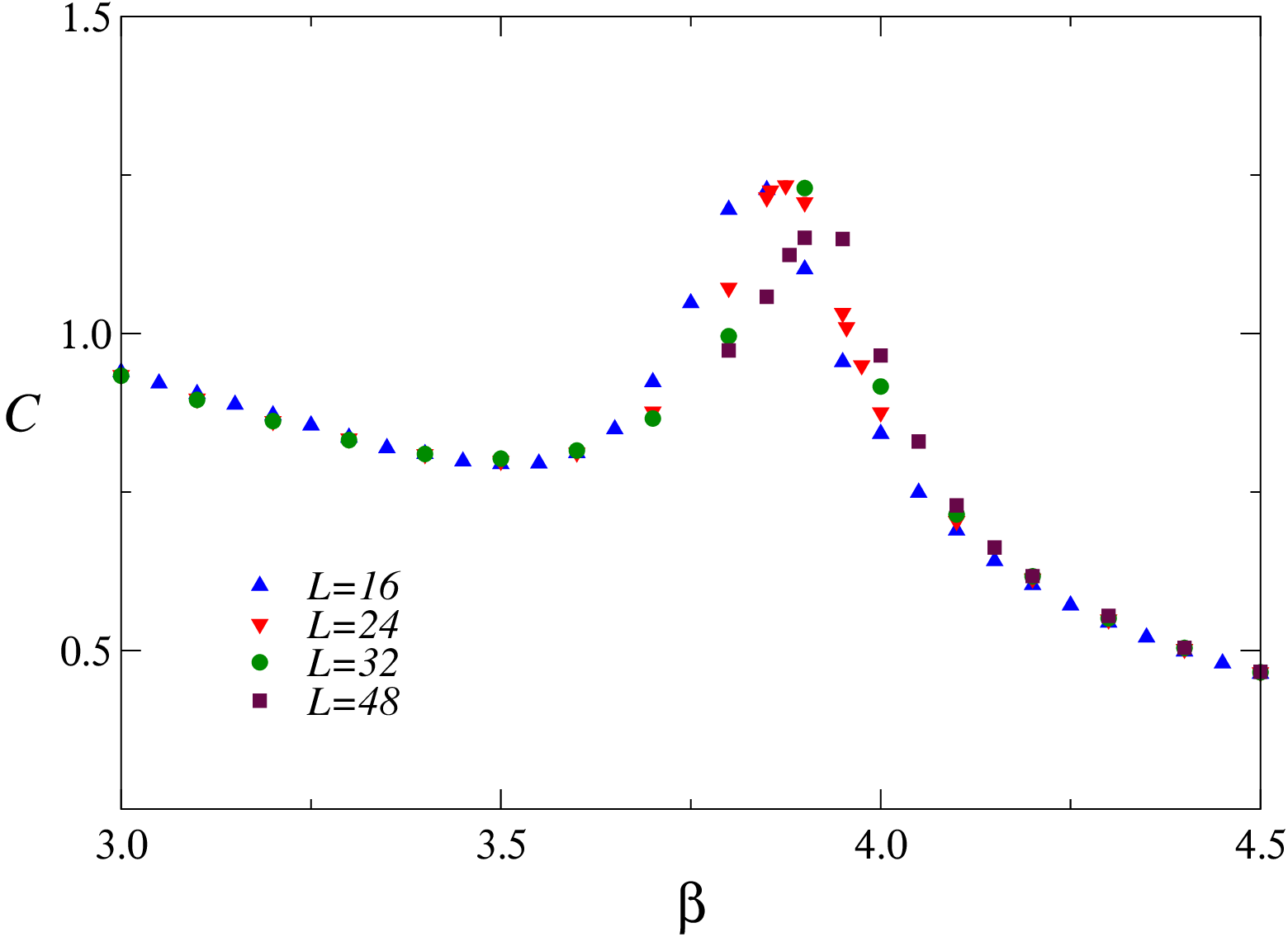}
  \includegraphics*[width=0.75\textwidth]{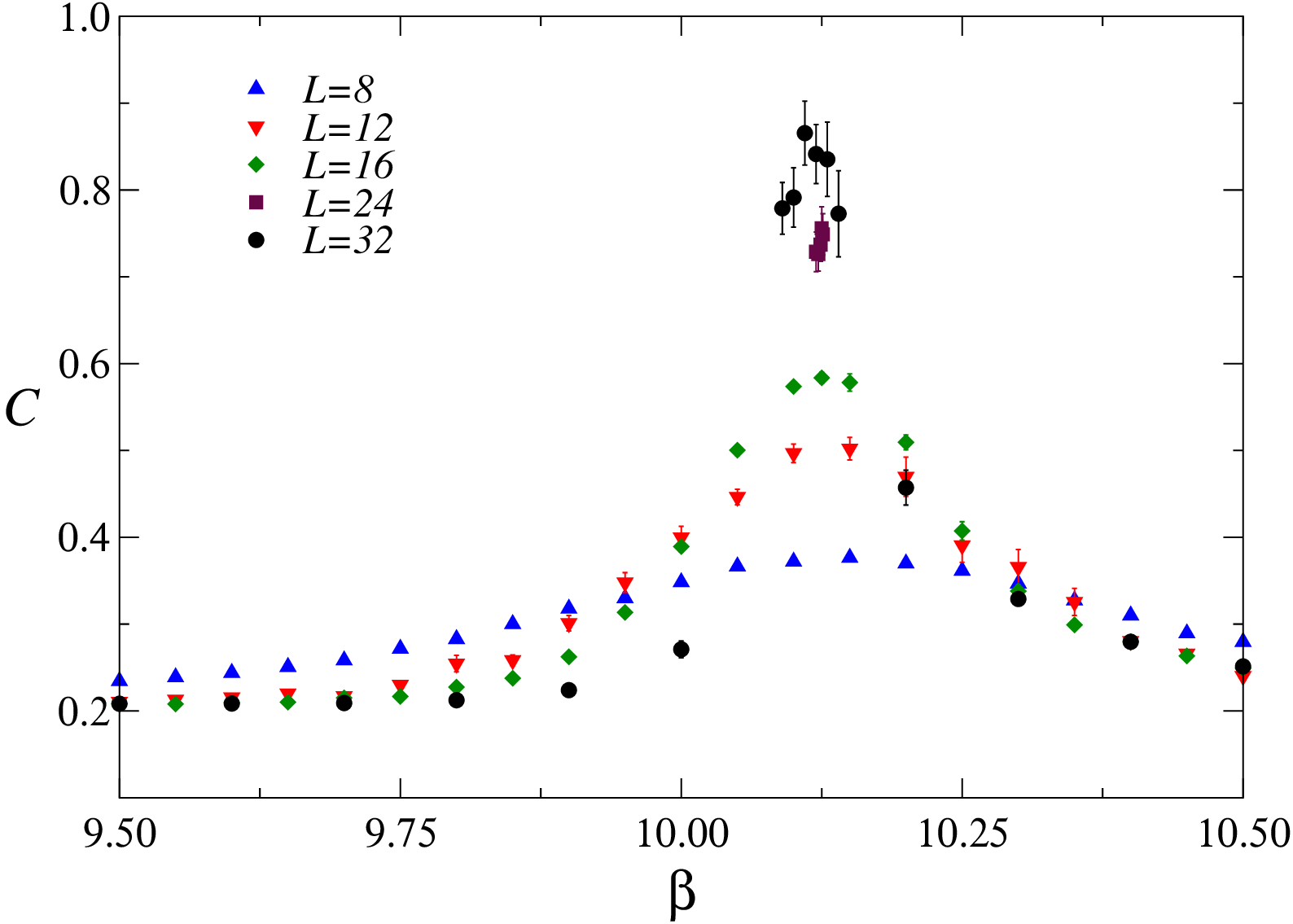}
  \caption{Plot of the specific heat $C$ versus $\beta$ for $N_c=2$,
  $N_f=4$ (top) and $N_c=3$, $N_f=3$ (bottom). In both cases 
  $\gamma=0$ and $v=1$.   
}
\label{crossover-specheat}
\end{figure}

\begin{figure}
  \centering
  \includegraphics*[width=0.75\textwidth]{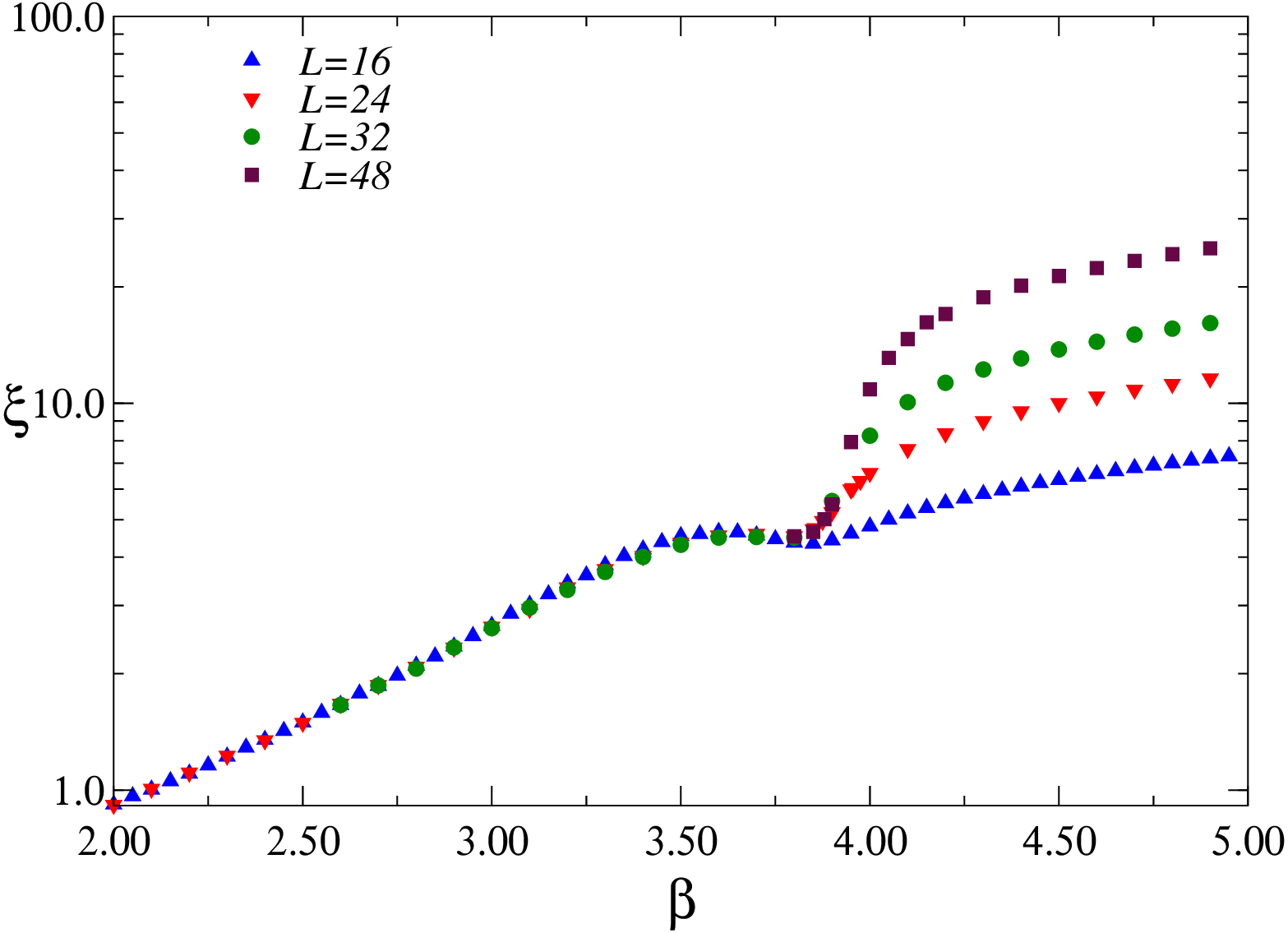}
  \caption{Plot of the $\xi$ versus $\beta$ for $N_c=2$,
  $N_f=4$, $\gamma=0$, and $v=1$. 
}
\label{crossover-xi}
\end{figure}

\subsection{Crossover from the $v=0$ to large-$v$ regimes}
\label{cross}

In Sec.~\ref{varsce} we have shown that, at $T=0$, the relevant configurations
depend on the sign of $v$, so that $v=0$ is a singular point for the 
large-scale behavior. The presence of this singularity at $\beta =\infty$ may,
in principle, give rise to singularities and discontinuities also at finite
$\beta$. For instance, there might be a first-order transition line 
$\beta = f_{FO}(v)$ in the $(\beta,v)$ plane that starts at $\beta=\infty$,
$v=0$ and ends at a finite value of $\beta$, with a critical endpoint where 
an Ising critical behavior is realized. Alternatively, it is possible that,
for finite $\beta$ only crossover phenomena occur without transitions.
We do not have performed a thorough analysis of this issue. However, in the 
few cases we have considered and that 
we discuss below, we have no evidence of finite-$\beta$ transition lines 
but only of crossover effects. 

In the simulations at fixed $v>0$ we have observed that the specific heat,
defined as  $C=\frac{1}{V}(\langle S^2\rangle -\langle S\rangle^2)$, has 
a strongly nonmonotonic behavior as $\beta$ increases, 
with a maximum at a finite value 
$\beta_{\rm max}$ of the inverse temperature. Two examples, corresponding to 
$N_c=2$, $N_f=4$ and $N_c=N_f=3$, are reported in  
Fig.~\ref{crossover-specheat}. At a first-order transition,
the maximum of the specific heat should increase with 
the lattice size, as $L^2$ in two dimensions.
In both cases shown in Fig.~\ref{crossover-specheat}, 
this does not occur: the maximum
of $C$ approaches a constant as $L$ increases, indicating that only a crossover
with no infinite-volume singularity
occurs. The crossover is obviously present in all quantities. In 
Fig.~\ref{nf4nc2v10xi} we have already reported 
the correlation length for $N_c=N_f=3$. It 
varies quite abruptly for $\beta\approx \beta_{\rm max}$.
The crossover is more evident in the behavior of $\xi$ for 
$N_c=2$, $N_f=4$, see Fig.~\ref{crossover-xi}. At $\beta \approx \beta_{\rm
max} \approx 3.8$, the behavior of $\xi$ changes abruptly, providing
a clear indication of a sudden change of the nature of the relevant
configurations. The presence of this crossover region  for $v \lesssim 1$, 
makes it difficult to determine the asymptotic behavior of the model for these
values of $v$. This is the reason why, in
Sec.~\ref{vg1nfl},  we considered the model at $v = 10$, a value of $v$ 
that is very far from the crossover region.

\section{Conclusions}
\label{conclu}

We have considered a class of 2D lattice non-Abelian gauge models with
$N_f$ scalar fields in the adjoint representation. They are defined by
the action (\ref{hgauge}) and are invariant under global O($N_f$) and
local SU($N_c$) transformations. For $N_f\ge 3$, the global symmetry
is nonabelian and thus, a critical (continuum) limit is only possible
in the limit $T\to 0$. We have therefore investigated the
zero-temperature behavior, to understand whether a continuum limit
exists and, if it does, to identify the corresponding 2D quantum field
theory. This work extend previous
results~\cite{BPV-19-ah2,BPV-20-qcd2,BFPV-20-ong,BFPV-21}, discussing
the role played by the gauge representations and by the quartic scalar
potential. For this purpose, we have identified the low-energy
configurations, that are relevant in the zero-temperature limit, and
we have derived effective models that are expected to describe the
large-scale behavior of the system.  The predictions have then been
checked numerically. We have performed MC simulations and determined
the universal features of the low-temperature behavior using FSS
methods.

We find that the continuum limit depends on the sign of the parameter
$v$ appearing in the scalar potential, see Eq.~(\ref{potential}).  For
$v\le 0$ the lattice gauge model has the same continuum limit as the
RP$^{N_f-1}$ model, for any value of $N_c$.  For positive $v$ instead,
the critical behavior depends on both $N_c$ and $N_f$.  For $N_f\le
N_c^2-1$, there is no continuum limit: correlation functions are
always short ranged.  On the other hand, for $N_f>N_c^2-1$ there are
long-range correlations for $T\to 0$. The corresponding continuum
limit is the same as that of the $\sigma$ model defined on the
symmetric space O($N_f$)/O($q$)$\otimes$O($N_f-q$) with $q=N_c^2-1$.
In particular, for $N_f=N_c^2$, the gauge model is equivalent to the
O($N_f$) vector $\sigma$ model.  Numerical data support these
predictions.  In particular, a FSS analysis of the MC data for $N_f=4$
and $N_c=2$ at $v=10$ clearly supports the prediction that the critical
behavior belongs to the universality class of the O(4) $\sigma$ model.

The results of this work provide additional support to the conjecture
that the critical behavior of any 2D lattice gauge model, defined
using the Wilson approach~\cite{Wilson-74}, belongs to the
universality class of a field theory associated with one of the
symmetric spaces that have the same global symmetry.

We finally mention that it is worth extending this study to analogous
three-dimensional systems, whose phase diagram is expected to be more
complicated, presenting various phases associated with the different
Higgs mechanisms that can be realized~\cite{SPSS-20,SSST-19}.  The
nature of transition lines separating the various phases are expected
to be crucially related to the interplay between global and local
gauge symmetries~\cite{BPV-19-sqcd}.

\appendix

\section{Role of the gauge fields for the minima}  \label{AppA}

In Sec.~\ref{varsce} we showed that, in the absence of gauge fields, there are
two minima whose relevance depends on the sign of the coupling $v$. We wish now 
to include the effects of the gauge fields. As suggested in 
Refs.~\cite{SSST-19,SPSS-20}, in the absence of the plaquette term, i.e., 
for $\gamma = 0$, we can integrate out the gauge fields, defining a local
effective potential: 
\begin{equation}
e^{-\beta \tilde{V}(D_{{\bm x},\mu})} = 
   \int d\widetilde{U} \, \exp[\hbox{Tr}\, 
     (\widetilde{U} D_{{\bm x},\mu})] \qquad 
    D_{{\bm x},\mu}^{ab} = 
   {\beta N_f\over 2} \sum_f \Phi^{af}_{{\bm x}+\hat{\mu}}  \Phi^{bf}_{\bm x}.
\end{equation}
If we assume translation invariance (and therefore drop the link dependence) 
and parametrize the field $\Phi^{af}$ as 
in Eq.~(\ref{singdec}), the matrix $D$ is given by
\begin{equation}
D = {\beta N_f\over 2} C W W^t C^t. 
\end{equation}
We are now interested in computing the integral in the limit $\beta\to \infty$
with the purpose of understanding whether the effective term $\widetilde{V}$
changes the conclusions obtained considering only $V(\Phi)$.
Note that, for $N_c = 2$, $\widetilde{U}$ is a generic orthogonal 
matrix and therefore we can easily check that the effective potential is 
independent of $C$ (it is enough to perform the change of variable
$\widetilde{U}' = C^t U C$). Such an independence is not a priori expected 
for $N_c > 2$ and thus the integral might depend both on $C$ and $W W^t$.

The integral reported here has been the object of several investigations, but
there is at present no exact general result, except for $N_c=2$, 
where one can take advantage of the results for the $O(N)$ link integrals 
\cite{BR-84,ESS-81} (some results for specific matrices $D$ are reported in 
Refs.~\cite{IZ-80,BM-84}). For $N_c = 2$ the result only depends on the 
eigenvalues of the matrix $W W^t$. If $N_f \ge 3$, these 
eigenvalues  coincide with $w_1^2$, $w_2^2$ and $w_3^2$, while for $N_f = 2$
one should consider $w_1^2$, $w_2^2$ and 0. The results reported in 
Ref.~\cite{BR-84} allow us to obtain 
\begin{equation}
\widetilde{V} = - {N_f\over2} (w_1^2 + w_2^2 + w_3^2) + {3\over 2 \beta}  
    \ln {\beta N_f\over2} + 
   {1\over2\beta} \ln [(w_1^2 + w_2^2) (w_1^2 + w_3^2) (w_2^2 + w_3^2)] +
      O(\beta^{-2}), 
\end{equation}
provided that at least two eigenvalues are not zero. If only one eigenvalue
is different from zero, one obtains \cite{BR-84}
\begin{equation}
\widetilde{V} = -{N_f\over2} w_1^2 + 
   {1 \over \beta} \ln {\beta N_f\over2} + {1\over \beta} \ln w_1^2 
    + O(\beta^{-2}).
\end{equation}
Since $w_1^2 + w_2^2 + w_3^2 = 2$ as a consequence of the constraint
$\hbox{Tr}\, \Phi^t \Phi = 2$, the leading contribution for $\beta \to \infty$
is independent of the field configuration. Thus, the gauge fields do not change
the conclusions on the relevant minimum configurations for $v>0$ and $v< 0$.
The calculation, however, allows us to determine the expected behavior for $v =
0$. Indeed, for the solution of type (I) (see Eq.~\ref{soluzioni}), the
subleading correction in  $\widetilde{V}$ is $\ln \beta/\beta$, which is
smaller than the the subleading correction, ${3\over 2} \ln \beta/\beta$, that
appears in $\widetilde{V}$ for configurations of type (II), for which 
$w_1^2 = w_2^2 = w_3^2 = 2/3$ (this is the relevant case for $N_f \ge 3$) or
$w_1^2 = w_2^2 = 1$, $w_3^2 = 0$ (this is the relevant case for $N_f=2$). 
This implies that, for $v = 0$, the asymptotic behavior is the same 
as for $v< 0$, i.e., in the RP$^{N_f}$ universality class. 

Let us now consider the case $N_c > 2$. In this case there is no general
formula for the integral. We will therefore assume that the relevant minima
are those that we have determined in Sec.~\ref{varsce} and, for each of them, 
we will determine the asymptotic behavior of the one-link integral.

We start by considering $(W W^t)^{ab}  = 2 \delta_{a1} \delta_{b1}$,
i.e. type (I) configurations.  If we set $v^a = C^{a1}$ (since $C$ is
orthogonal, $v^a$ is a unit vector), we have
\begin{equation}
  D^{ab} = \beta N_f v^a v^b
\end{equation}
The integral can then be written as 
\begin{equation}
e^{-\beta \widetilde{V}} = \int dU \, 
   \exp[\hbox{Tr} (U^\dagger M U M) ] 
\qquad M = (2 \beta N_f)^{1/2} \sum_a v_a T^a \;. 
\label{integral-IZ}
\end{equation}
The matrix $M$ is hermitean and traceless. If $\lambda_a$ are its eigenvalues, 
we have 
\begin{equation}
\sum_a \lambda^2_a = \hbox{Tr} \, M^2 = \beta N_f \sum_a v_a^2 = 
\beta N_f.
\end{equation} 
Integral (\ref{integral-IZ}) has been computed in Ref.~\cite{IZ-80}. The
leading term can be rewritten in terms of the determinant of the matrix
$\Lambda$, whose elements are $\Lambda_{ab} = e^{\lambda_a \lambda_b}$:
\begin{equation}
\widetilde{V} = -\frac{\ln\det \Lambda}{\beta} + O(\ln \beta/\beta) \;. 
\end{equation}
Now, $\det \Lambda$ is a sum of terms of the form 
\begin{equation}
\exp[\lambda_1 \lambda_{i_1} + \lambda_2 \lambda_{i_2} + 
    \ldots + \lambda_N \lambda_{i_N}], 
\end{equation}
where $(i_1,\ldots, i_N)$ is a permutation of $(1,\ldots N=N^2_c - 1)$
As a consequence of the Schwartz inequality 
\begin{equation}
\lambda_1 \lambda_{i_1} + \lambda_2 \lambda_{i_2} + 
    \ldots + \lambda_N \lambda_{i_N} \le  
\sum_a \lambda_a^2 = \beta N_f, 
\end{equation}
the equality being obtained for $i_1 = 1$, $i_2 = 2$, $\ldots$ $i_N = N$.
This implies that $\det \Lambda \sim e^{\beta N_f}$ for $\beta \to \infty$. 
We thus obtain for $N_c > 2$
\begin{equation}
\widetilde{V} = - N_f + O(\ln \beta/\beta).
\label{largeV}
\end{equation}
Let us now consider the configurations of type (II). We have not been able to
obtain results for $N_f < N_c^2 - 1$, in which $WW^t$ is a diagonal matrix
that has both zero and unit eigenvalues. The case
$N_f \ge N_c^2 - 1$, is instead easily discussed. The matrix $W W^t$ is 
proportional to ${1\over q} I$ and thus we obtain
\begin{eqnarray}
e^{-\beta\widetilde{V}} &=& 
   \int d\widetilde{U} \, 
    \exp\left( {\beta N_f\over q} \hbox{Tr}\ \widetilde{U} \right) 
    = e^{-\beta N_f/q} \int dU \,  
     \exp\left( {\beta N_f\over q} |\hbox{Tr}\ {U}|^2 \right) .
\end{eqnarray}
The large $\beta$ behavior of the integral is obtained 
by expanding around $U = I$ (a few terms of the expansion are 
obtained in Ref.~\cite{BM-84}). The leading term for the integral is 
$\exp(\beta N_f N_c^2/ q)$, which gives again Eq.~(\ref{largeV}). As it happens
for $N_c=2$, the gauge contribution to the potential is the same 
for both types of minimum configurations.

To conclude the Appendix, let us note that the gauge effective potential 
$\widetilde{V}$ would play a different role if one considers a different 
approach to $T=0$. Indeed, let us define (again for $\gamma=0$)
\begin{equation}
Z = \sum_{\{\Phi, U\}}  \, 
  \exp[-\beta S_K(U) + f(\beta) S_V(\Phi)]
\end{equation}
where $f(\beta)$ is a function of $\beta$. In our work we have considered 
$f(\beta) = \beta$, but one can also consider functions with 
a different large-$\beta$ behavior. One possibility consists in taking 
$f(\beta)$ finite for $\beta\to 0$. In this case the potential would play no 
role and the dominant term would be the gauge potential $\widetilde{V}$.
Therefore, the critical behavior would be independent of $v$, the same as 
that we observe for $v=0$. A nontrivial behavior would only be obtained 
by selecting a function $f(\beta)$ that behaves as $\ln \beta$ as 
$\beta\to\infty$. In this case both the gauge contribution and the 
scalar potential would play a role. One would expect a critical $v_c$,
such that different phases are realized for $v > v_c$ and $v < v_c$.

\section{Monte Carlo simulations: technical details}
\label{MCsim}

We performed MC simulations on square lattices with periodic boundary
conditions.  The gauge link variables $U$ were updated using a standard
Metropolis algorithm \cite{Metropolis:1953am}.  The new link variable was
chosen close to the old one, in order to guarantee an acceptance rate of
approximately 30\%. The scalar fields were updated using two different
Metropolis algorithms, again tuning the proposal to obtain an acceptance rate
of 30\%. The first update performs a rotation in flavor space
\begin{equation}
    \phi^f \mapsto (O\phi)^f, \quad O \in \text{SO}(N_f),
\end{equation}
while the second one rotates the colors of a single flavor
\begin{equation}
    \phi^f \mapsto (H\phi H^\dagger)^f, \quad H \in \text{SU}(N_c).
\end{equation}
In the simulations with $v=0$, since the action is linear in the scalar fields,
we also considered microcanonical steps \cite{Creutz:1987xi} implemented
\emph{\`a la} Cabibbo-Marinari \cite{Cabibbo:1982zn} (the relative frequency of
Metropolis and microcanonical updates was chosen equal to 3/7). Microcanonical
updates could not be used for $v\neq 0$, since the action is not linear in the
scalar fields.  
Typical statistics of our runs, for a given value of the parameters
and of the size of the lattice, were of order of 10$^7$-10$^8$ lattice sweeps
(in a sweep we update all lattice variables once), with the largest number associated to 
runs performed without the microcanonical update.
Errors were estimated using a standard blocking and jackknife
procedure, with a maximum blocking size of the order of $10^5$ updates.

\bigskip

\acknowledgments

Numerical simulations have been performed on
the CSN4 cluster of the Scientific Computing Center at INFN-PISA.

\end{document}